\documentclass[paper]{ieice}

\normalsize

\DeclareMathAlphabet\mathbfcal{OMS}{cmsy}{b}{n}

\usepackage{authblk}
\usepackage{algorithm}
\usepackage{algorithmic}
\usepackage{graphicx,amsmath,amssymb,latexsym,epsfig}
\usepackage{setspace}
\usepackage{psfrag}
\usepackage{array,booktabs,arydshln,xcolor}

\usepackage{lscape}
\newtheorem{problem}{Problem}

\field{B}
\begin{document}


\title[Kalman Prediction Based Proportional Fair Resource Allocation]{Kalman Prediction Based Proportional Fair Resource Allocation for a Solar Powered Wireless Downlink}

\authorlist{%
\authorentry{Neyre Tekbiyik-Ersoy}{}{labelA}
\authorentry{Elif T. Ceran}{}{labelB}
\authorentry{Kemal Leblebicioglu}{}{labelC}
\authorentry{Tolga Girici}{}{labelD}
\authorentry{Elif Uysal-Biyikoglu}{}{labelE}
}
\breakauthorline{2}
\affiliate[labelA]
{Cyprus International University}
\affiliate[labelB]
{Imperial College London}
\affiliate[labelC]
{Middle East Technical University}
\affiliate[labelD]
{TOBB University of Economics and Technology}
\affiliate[labelE]
{Ohio State University}

\bibliographystyle{ieicetr}

\maketitle

\def\eg{\emph{e.g.}}
\def\ie{\emph{i.e.}}

\begin{summary}
Optimization of a Wireless Sensor Network  (WSN) downlink with an energy harvesting transmitter (base station) is considered. The base station (BS), which is attached to the central controller of the network, sends control information to the gateways of individual WSNs in the downlink. This paper specifically addresses the case where the BS is supplied with solar energy. Leveraging the daily periodicity inherent in solar energy harvesting, the schedule for delivery of maintenance messages from the BS to the nodes of a distributed network is optimized. Differences in channel gain from the BS to sensor nodes make it a challenge to provide service to each of them while efficiently spending the harvested energy. Based on PTF (Power-Time-Fair), a close-to-optimal solution for fair allocation of harvested energy in a wireless downlink proposed in previous work, we develop an online algorithm, PTF-On, that operates two algorithms in tandem: A prediction algorithm based on a Kalman filter that operates on solar irradiation measurements, and a modified version of PTF. PTF-On can predict the energy arrival profile throughout the day and schedule transmission to  nodes to maximize total throughput in a proportionally fair way.

\end{summary}

\begin{keywords}
Broadcast, energy harvesting, proportional fairness, time sharing, industrial wireless sensor networks, solar energy, Kalman filter, prediction.
\end{keywords}

\section{Introduction}

Due to their rapid deployment, flexibility, as well as collaborative sensing, self-organization,  and intelligent-processing abilities, wireless sensor networks (WSNs) have found their way into a wide variety of industrial settings with varying requirements and characteristics: healthcare monitoring~\cite{AlemdarC10}, structural health monitoring, pipeline monitoring~\cite{YoonYHLS11}, agricultural monitoring~\cite{Riquelme09}, networked
control systems (NCSs) where a spatially distributed feedback control system is required \cite{Ulusoy2011}, etc. In particular, in industrial WSN applications, an area needs to be covered with one or multiple WSNs, monitoring different parameters, or different locations. Often, these subnetworks of simple devices send data via gateway nodes (or cluster heads) to a remote BS located at a central office, where the signal processing to produce strategic decisions runs on a more powerful computer. It is then necessary for the BS to regularly broadcast certain network details and commands to the nodes. Sustainable and environmentally friendly development of such industrial applications requires increased use of renewable energy, e.g., solar or wind power.

Resource management is very critical in the industrial environment. Since industrial WSNs are expected to be deployed in harsh or inaccessible environments for long periods of time, a remote BS may be needed to control the operation of these networks. Recently, employing energy harvesting (via ambient energy sources such as solar irradiation~\cite{Bogue12}, vibrations~\cite{MohamedWM11}, and wind~\cite{YenP11}) to power transmitters of wireless networks, such as BSs has gained tremendous interest. Today, solar energy is becoming widely used, due to its high power density compared to other sources of ambient energy~\cite{NohK09}. Therefore, the scenario considered in this paper involves an outdoor industrial WSN (or a combination of WSNs) controlled by a central node that is powered by solar panels, at least as an auxiliary energy source. During times of low or no daylight, the BS will need to rely on another power source. To reduce the dependence on these other sources and maximize efficient use of solar power, it makes sense to schedule time-insensitive communications from the BS to the nodes to times of high solar irradiation. 

Of course, outdoor solar irradiation exhibits a daily periodicity. However, there are seasonal as well as short-term variations. Depending on such a varying energy source will require the revision of conventional resource management. When, for example, the size of a solar cell limits the available power, decisions about when to provide how much power, rate, service, etc. have to be made. As also stated in~\cite{Koksal2012}, conservative energy expenditure, may lead to missed recharging opportunities if the battery is already full. On the other hand, aggressive usage of energy may result in reduced coverage or connectivity for certain time periods, that could make the BS temporarily incapable of transferring time-sensitive data. In industrial applications, this may lead to loss of production and may sometimes create hazardous situations. Hence, new resource allocation and scheduling schemes need to be developed to balance these contradictory goals, in order to maximize the network performance. 

In this paper, we focus on a scenario of a solar powered BS that needs to send protocol maintenance messages to sensor nodes (or, to a set of gateways or cluster heads). It is well known that in WSNs, maintenance messages (topology updates, protocol information, etc) constitute a significant fraction of all messages that are passed. If the BS needs to deliver these routine messages to the nodes, how should it schedule these over the duration of a day, to maximize its efficient use of solar energy? Furthermore, how much power and time should it allocate to different nodes which, being spread over an area, may observe vastly different path losses from each other? 
If the problem was formulated as a throughput maximization problem, the solution would essentially constitute of always sending to the node that is closest to the transmitter, which defeats the purpose of serving all the gateways. Hence, it is more appropriate to construct a formulation that maximizes throughput in a \textit{proportionally fair} way so that no particular gateway is starved due to its high path loss from the transmitter. The contribution of this paper is to combine a Kalman-filter based solar energy prediction algorithm with a proportional-fair scheduler. 
The proportional-fair energy harvesting resource allocation problem formulated in~\cite{6463487} was shown to be a {\emph{biconvex}} problem\footnote{The problem of optimizing a biconvex function over a given (bi)convex or compact set, where a function $f : X\times{Y}\rightarrow{\Re}$ is called biconvex if $f(x,y)$ is convex in $y$ for fixed $x\in{X}$ and is convex in $x$ for fixed $y\in{Y}$~\cite{Pirsiavash10}. } which is nonconvex, and has multiple optima. The optimum off-line schedule developed in~\cite{tekbiyik2} (which assumes that the energy arrival profile at the transmitter is deterministic and known ahead of time in an off-line manner), a Block Coordinate Descent based optimization algorithm, BCD, was shown to converge to an optimal solution for proportional fair allocation of harvested energy in a wireless downlink. A simple heuristic, called PTF, that can closely track the performance of the BCD solution was also developed in~\cite{6463487}. However, in many practical scenarios, the energy harvests are not known a priori. Thus, this paper leverages the PTF algorithm to develop an \textit{online} resource allocation algorithm, PTF-On. PTF-On is a stand-alone algorithm that operates two algorithms in tandem: A Kalman filter-based solar energy prediction algorithm, and a modified version of the PTF algorithm. PTF-On can predict the BS's energy arrival profile throughout the day, and then, act upon this energy arrival profile to determine the best power and time allocation that will maximize the throughput (the amount of data sent to the gateways) in a proportionally fair way. 

We start by summarizing some related work in the next section and continue by describing our system model in Section \ref{sec:sm}. After that, in Section \ref{sec:psandst}, we explore the structure of the problem. Our Kalman-based solar prediction algorithm is described in Section \ref{sec:k-sep}. Section \ref{sec:ptf-on} proposes the online allocation algorithm, PTF-On. In Section \ref{sec:theresultsofchp7}, we test our proposed algorithms with respect to several related schemes. We conclude in Section \ref{sec:conc} with an outline of further directions.


\section{Related Work}
\label{sec:rw}
Recently, several works in the field of Industrial WSNs have been conducted.  In \cite{Manfredi2010}, a sink resource allocation strategy based on log-utility fairness criteria is proposed. In \cite{Zhang2013}, authors propose a solution to the problem of maximizing the minimum energy reserve. An algorithm to achieve arbitrarily close to optimal power efficiency while satisfying the desired estimation accuracy of processes over some time is presented in \cite{Vijayandran2011}. Authors have expanded their work by introducing a WSN model that includes fairness control among sensors and both energy harvesting batteries and backup battery units.  An optimal resource allocation presented in the paper is used for state estimation to provide a desired accuracy constraint by taking advantage of Kalman filter. Even if proposed WSN frame includes energy harvesting batteries modelled as IID random processes, the renewable solar energy may not be modelled as an IID process. Moreover, Kalman filter approach used in \cite{Vijayandran2011} is used to estimate the state of the processes to be detected by sensors, not to predict solar irradiations as this paper presents. 

In \cite{Sharma2010}, Sharma et.al. present throughput optimal energy management policies for energy harvesting sensor networks. However, results are valid only for stationary ergodic energy harvesting and data processes. Outdoor solar irradiation, which has a general 24-hour variation cannot be well modelled as a stationary process. On the other hand, Sharma et. al also propose optimal energy management policies for energy harvesting sensor nodes with solar energy harvesting. While solar irradiation cannot be considered as a stationary ergodic process, it is assumed to be piecewise stationary over half an hour periods  \cite{Sharma}.  Differently from the work in \cite{Sharma}, the problem in our case has only sub-optimal solutions as proved in \cite{6463487} and prediction intervals for solar energy harvesting are applied within periods of half an hour. 
In \cite{Lalitha2013}, the problem of minimizing the average grid power consumption of a Green BS downlink in scheduling N users with average delay constraints is considered and formulated as Markov Decision problem. Any fairness criteria is not regarded in the work and harvested energy is taken as a stationary IID process without taking into account the statistics and daily periodicity of the solar irradiation. 
In \cite{Lin2007}, authors present an asymptotically optimal energy aware routing model for WSNs and propose an online algorithm which does not know future packet requests. However, as in most of the similar works, the energy model assumes that short term energy replenishment schedule for nodes is known. While there are studies on prediction of solar irradiation such as \cite{Chaabene2008}, \cite{Lonij2013} and \cite{Hassanzadeh}; few have  used Kalman filtering techniques, and none of them combines  online proportional time fair resource allocation algorithm  with such a predictor.

To the best of our knowledge, there is no prior work in the literature  about proportional fair resource allocation in  WSNs with a prediction algorithm based on a Kalman filter that operates on solar irradiation measurements.  We propose an online method for scheduling nodes in an industrial WSN according to predicted solar energy levels,  with a fair resource allocation criterion.

\section{System Model}
\label{sec:sm}
\begin{figure}[tb]
	\centering
		\includegraphics[width=0.45\textwidth, height=0.7\textwidth]{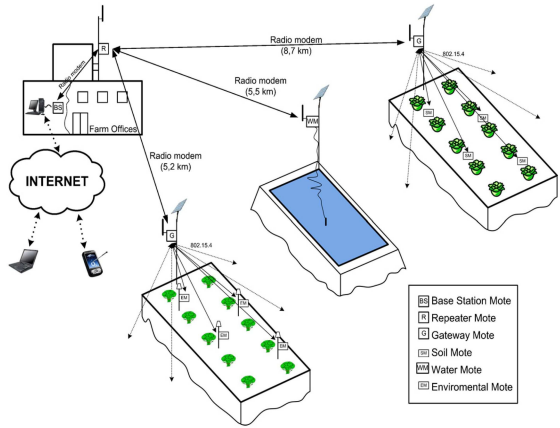}
		\centering
		\vspace{-2.4in}
\caption{Example industrial WSN application (agricultural monitoring) controlled by a remote base station ~\cite{Riquelme09}.} 
\label{fig:fig2}		
\end{figure}

\begin{figure*}[tb]
	\centering
		\includegraphics[width=1\textwidth, height=0.15\textwidth]{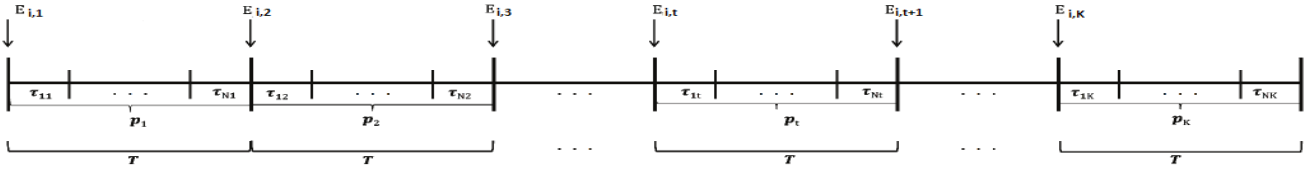}
		\vspace*{-0.22in}
\caption{One of the multiple frames in a timeline. The highlighted frame $i$ (of 24 hours) includes $K$ energy arrivals. The time between consecutive arrivals is allocated to $N$ users.} 
\label{fig:problemillustration1}		
\end{figure*}

  The goal is to schedule transmissions from a BS to $N$ gateways (or cluster heads, or individual sensor nodes, without loss of generality), over a certain time window that we refer to as a \textit{frame}. The transmissions to different receivers will be organized by time sharing over a bandwith W. The setup is described in Figure \ref{fig:fig2}. We assume arbitrary channel gains $g_n$ from the BS to receiver $n$, $n\geq 1$, and, w.l.o.g, for simplicity, constant noise Power Spectral Density $N_o$ at all receivers. The channel gains are thought of as long term average gains, when short term channel variations are averaged out. Hence, $g_n$'s are constant throughout the frame and are known by the scheduler. 
The BS is equipped with a rechargeable battery, powered by a solar panel, such that harvested energy becomes available at distinct instances. The durations between two harvest instants will be called a ``slot'' (as in~\cite{6463487} ). Our system model is based on the one illustrated in Figure \ref{fig:problemillustration1}. 
For facilitating daily predictions, we set the length of a frame to 24 hours. Note that, we restrict our attention to the case of periodic energy arrivals ($T_t=T$ for all $t\in \left\{1,\hdots,K\right\}$), as in~\cite{6463487}). Not all generality is lost, since harvest amounts are arbitrary and the absence of a harvest in a certain duration can be expressed with a harvest of amount zero for the respective slot.  The amount of energy harvested from the environment at the beginning of time slot $t$ of frame $i$ is $E_{i,t}$. The BS chooses a power level $p_t$ and a time allocation vector $\tau_t=(\tau_{1t},...,\tau_{Nt})$, for each time slot $t$ of the frame, where $p_{nt}=p_t$ is the transmission power for gateway $n$ during slot $t$ and, $\tau_{nt}$ is the time allocated for transmission to gateway $n$ during slot $t$.


\section{Problem Statement and Structure}
\label{sec:psandst}
We define the total achievable rate for user $n$ (the number of bits transmitted to user $n$) within a frame, $R_n$. Our goal is to maximize a total utility, i.e., the log-sum of the user rates $\sum_{n=1}^Nlog_2(R_n)$, which is known to result in proportional fairness~\cite{MaoKS10}. Without loss of all generality\footnote{Most practical rate-power relationships will satisfy convexity and may be used in a similar formulation. Our choice of rate function here following AWGN capacity is quite standard.}, by using AWGN (Additive Gaussian Noise) channel capacity as a rate function to construct a biconvex problem \cite{6463487}, we define $R_n=\sum_{t=1}^K\tau_{nt}W\log_2\left(1+\frac{p_{t}g_n}{N_oW}\right)$. Thus, we obtain the constrained optimization problem, Problem \ref{pr:DownlinkScheduling}, where (\ref{eq:nonnegativity2}) represents the nonnegativity constraints for $t=1,...,K$ , $n=1,...,N$. The equations in (\ref{eq:Timeconstraint}), called time constraints, ensure that the total time allocated to users does not exceed the slot length, and, every user gets a non-zero time allocation during the frame. Finally, the equations in (\ref{eq:Energycausality}), called energy causality constraints, ensure no energy is consumed before becoming available. 
\begin{problem}
\label{pr:DownlinkScheduling}
\small \begin{align}
\noindent\mbox{Maximize:  } &U(\overline{\tau},\overline{p})=\sum_{n=1}^N\log_2\left(\sum_{t=1}^K\tau_{nt}W\log_2\left(1+\frac{g_np_{t}}{N_oW}\right)\right)\nonumber \\
\noindent \mbox{subject to: }&\tau_{nt}\geq{0}  \ , \ p_{t}\geq{0} \label{eq:nonnegativity2} \\ 
&\sum_{n=1}^N\tau_{nt} = T_t \label{eq:Timeconstraint} \ , \ \sum_{t=1}^K\tau_{nt} \geq \epsilon \\
&\sum_{i=1}^tp_{i}T_{i} \leq{\sum_{i=1}^tE_i} \label{eq:Energycausality}
\end{align}\normalsize
\end{problem} Note that Problem \ref{pr:DownlinkScheduling} is a biconvex optimization problem with multiple optima, and there exists an offline heuristic algorithm, PTF~\cite{6463487}, that can closely track the optimal solution (solution found by BCD~\cite{6463487}) of this problem. With the purpose of adapting real life scenarios, in this paper, we modify the PTF algorithm so that we can use it in an online setting, i.e., the amounts of energy harvests within a frame are not known a priori. The modified version of the PTF algorithm will need to be combined with an energy prediction algorithm, which will be explained in the next section.


\section{Kalman-Based Solar Energy Prediction}
\label{sec:k-sep}

In this section, we apply the Kalman filter algorithm to forecast the energy arrivals within a frame, for a BS powered with solar panel. We consider sub-hourly prediction of the energy arrivals for a frame of 24 hours (one day) as an example, and, formulate the Kalman filter for the following state and measurement models:
\begin{align}
x(k+1)&=\alpha_1x(k)+\alpha_2x(k-47)+\beta_1y(k)+w(k) \label{eq:state_model} \\
z(k)&=x(k) + v(k) \label{eq:meas_model}
\end{align} where $x$ and $z$ represent the state (energy level) and the measurement respectively. This model is mainly based on the idea that; due to the diurnal cycle of a day, the amount of energy that will be harvested in the $(k+1)^{th}$ sub-hour of an arbitrary day, $x(k+1)$, should be related to the energy harvested in the $k^{th}$ sub-hour of the same day, $x(k)$, the solar irradiation received in the $k^{th}$ sub-hour of the same day, $y(k)$, and, the energy harvested in the $(k+1)^{th}$ sub-hour of the previous day (the energy that was harvested 48 sub-hours ago: $x((k+1)-48)=x(k-47)$), $x(k-47)$. In (\ref{eq:state_model}), $w(k)$ is a modeling error, which represents the effects of the uncontrolled events on the harvested energy (such as shadowing caused by clouds passing through, disturbance to the solar panel, or damage due to malicious act, etc.). It is modelled as IID Gaussian with zero mean and variance $\sigma^2_w$. The parameters $\alpha_1$,$\alpha_2$ and $\beta_1$ represent the weights assigned to emphasize the importance of the parameters that will be used for prediction. In the measurement model, $v$ denotes the IID Gaussian measurement noise with zero mean and variance $\sigma^2_v$.
By considering that there are 48 sub-hours in a day, the overall state equations can be re-stated in matrix form as in (\ref{eq:stateeq}).
\begin{align}
\label{eq:stateeq}
\footnotesize
\begin{bmatrix}
x(k+1)\\
x(k)\\
x(k-1)\\
\vdots \\
x(k-46)
\end{bmatrix}
=
\textbf{A}
\begin{bmatrix}
x(k)\\
x(k-1)\\
x(k-2)\\
\vdots \\
x(k-47)
\end{bmatrix}
+\footnotesize
\overline{\beta}
y(k)+
\overline{\Gamma}
w(k)
\end{align} Now, we define an augmented state vector, $\overline{\xi_{k}}$, which contains the energy amounts harvested today:
\begin{align}
\overline{\xi_{k}}=
\begin{bmatrix}
x(k)&x(k-1)&\hdots&x(k-47)
\end{bmatrix}'
\end{align} We define a new matrix $\textbf{A}$, column vectors $\overline{B}$, and $\overline{\Gamma}$: \begin{align}
\textbf{A}&=
\footnotesize\begin{bmatrix}
  \alpha_1 & 0 & 0 & \hdots & 0 & 0 & \alpha_2\\
  1 & 0 & 0 & \hdots & 0 & 0 & 0  \\
  \vdots & & & \ddots & & & \vdots\\
   0 & 0 & 0 & \hdots & 0 & 1 & 0 \\  
\end{bmatrix} \\
\overline{B}&=
\footnotesize\begin{bmatrix}
\beta_1 & 0 & \hdots &0  
\end{bmatrix}'     
\hspace{0.2in}\overline{\Gamma}=\hspace{0.05in} \footnotesize\begin{bmatrix}
1 & 0 & \hdots  &0  
\end{bmatrix}'
\end{align} Thus, the state model in (\ref{eq:stateeq}), and the measurement model in (\ref{eq:meas_model}) reduce to:
\begin{align}
\overline{\xi_{k+1}}&=\textbf{A}\overline{\xi_{k}}+\overline{B}y(k)+\overline{\Gamma}w(k) \label{eq:truth1}\\
z(k)&=x(k) + v(k)
\label{eq:truth2}
\end{align} which is structurally equivalent to the ``truth'' model described in (5.27) of~\cite{Crassidis04}. Thus, by applying the Discrete-Time Linear Kalman Filter described in~\cite{Crassidis04}, we are able to predict the amount of energy arrival in the next sub-hour by only using the amount of energy arrival in this sub-hour, the solar irradiation received in this sub-hour and, the arrival in the previous day's next sub-hour.Please note that, in order to compute the best weights $\alpha_1$, $\alpha_2$ and $\beta_1$ that will be used for simulations, we use a data fitting method described as follows: By using the 18 days' data (real power measurements belonging to 01-18.10.2009 for Amherst, Massachusetts, USA) provided by Navin Sharma~\cite{SharmaGIS10}, we design a Newton algorithm that aims to minimize the Mean Squared Error (MSE) between the data obtained from real measurements and the estimated data according to the state and measurement models in (\ref{eq:state_model}) and (\ref{eq:meas_model}). Thus, the objective function  to be minimized by the Newton algorithm is described below:
\begin{align}
\frac{1}{N}\sum^{N}_{k=1}(z(k)-z_m(k))^2
\end{align} 
where $z$ denotes the data obtained from actual measurements and $z_m$ denotes the estimated data obtained from the models, in (\ref{eq:state_model}) and (\ref{eq:meas_model}). Note that we have $48$ subhours for a day, and need the past day's data at the same subhour for the prediction of a subhour's solar irradiation. For 17 days ($17$ days$=816$ sub-hours) data ~\cite{SharmaGIS10}, the objective function can be stated as: 
\begin{align}
\frac{1}{816}\sum^{863}_{k=48}(z(k+1)-(\alpha_1x(k)+\alpha_2x(k-47)+\beta_1y(k)))^2
\end{align}
Our simulation results, provided in Section \ref{sec:theresultsofchp7}, show that the best values for weights, $\alpha_1$,$\alpha_2$,$\beta_1$ are 0.7184, 0.1439, and, 0.0063 respectively, when the $x(k)$'s are in terms of kilojoules and  the initial values for the data fitting operation of $\alpha_1$,$\alpha_2$,$\beta_1$ are taken as 0.9, 0.1 and 0.01 respectively. Furthermore, considering the equivalence of the ''truth" model in \cite{Crassidis04}, the prediction and update equations of the Kalman estimator can be stated as:
\begin{align}
\overline{\xi_{k+1}^-}=\textbf{A}\overline{\xi_{k}^+}+\overline{B}y(k), 
\hspace{0.05in}\overline{\xi_{k}^+}=\overline{\xi_{k}^-}+\textbf{K}_k[z(k)-\overline{\xi_{k}^-}]
\end{align}
where $\xi_k^-$ and $\xi_k^+$ denotes the pre-measurement and post-measurement states respectively. Moreover, K, R , I, P are the Kalman gain function, measurement noise matrix, identity matrix and error covariance matrix respectively and defined as:
\vspace{-0.1in}
\begin{align}
K_k&=P_k^--[P_k^-+R]
\\
P_k^+&=[I-\textbf{K}_k]P_k^-
\\
P^-_{k+1}&=\textbf{A}P_k^+\textbf{A}^T+\overline{\Gamma}\sigma_\omega^2\overline{\Gamma}^T 
\end{align}Similar to the notation of states $\xi_k^-$ and $\xi_k^+$, $P_k^+$ and $P_k^-$ denote the pre-measurement and post measurement error covariance matrices. Thus, we have KSEP with the state and measurement models in (\ref{eq:state_model}) and (\ref{eq:meas_model}).


\section{PTF-On Algorithm}
\label{sec:ptf-on}

In this section, we propose an online proportional fair resource (power and time) allocation algorithm, called PTF-On. PTF-On is the online version of the PTF heuristic proposed in~\cite{6463487}. Note that the PTF algorithm operates in an offline fashion, i.e., the energy arrival amounts within a frame are known at the beginning of that frame. The main motivation of the PTF-On algorithm can be explained as follows: There are 48 sub-hours and thus, 48 energy arrivals within a frame (24 hours). At the beginning of a each slot, the current amount of residual energy and amounts of previous harvests are known. The amounts of next 47 energy arrivals should be predicted. Thus, at the beginning of each frame we perform two prediction operations to determine the energy amounts that will be harvested during the frame. We perform this operation as follows: At the beginning of Slot 1, the energy arrives and is known to the BS. Thus, the BS can use K-SEP (Kalman Based Solar Energy Prediction) to predict 
 its next energy arrival, i.e., the arrival in Slot 2.  The arrivals other then the arrival in the next slot can not be predicted before a sub-hour passes without enlarging the error covariance matrix . This is due to the fact that a half an hour should pass to see what is really harvested in Slot 2, so that this value can be used to predict and verify the value in Slot 3. In case of predicting the arrivals other than the first successive arrival, we need to deal with bigger covariance matrices. Thus,to be able to deal with simpler computations, we adopt S-SEP, which does not use the data that was harvested in the previous slots (mainly predicts the amount of energy that will be harvested in today's $k^{th}$ sub-hour as the average of the energy arrival amounts of the past two days' $k^{th}$ sub-hours), to predict the next 46 arrivals. Thus, all energies (or at least their estimates) are known to the BS at the beginning of the frame. This way, at the beginning of each frame, one can run PTF algorithm to determine a close-to-optimal power and time allocation that will maximize the throughput in a proportionally fair way, for the up-coming 24 hours.

PTF-On requires past two days' data for predicting the energy arrival amounts of the day it will be used in. Assume that there are days 1,2,3,4,... etc, and, PTF-On will be used to predict the arrivals, and, determine the most proportional fair resource allocation, for the second half of day 3 and first half of day 4 (Frame of 24 hours: From 12:00 of day 3 to 12:00 of day 4). The operation of PTF-On algorithm is explained below.
\begin{enumerate}
	\item For the 24-hours frame started at 12:00 of day 3, there will be 48 slots, each 30 minutes of length (Please note that this frame is called the original frame). The beginning of the whole frame will be the beginning of Slot 1. Thus, when the frame starts, the energy arrival at the beginning of slot 1 of day 3, $E_{3,1}$ is known. Thus, the energy arrival at the beginning of Slot 2, $E_{3,2}$, can be predicted by using the K-SEP algorithm. Then, use the S-SEP algorithm to obtain rough predictions of the others, $E_{3,3},\hdots,E_{3,48}$, and form a predicted harvest series as follows: $\overline{E_{pred}}=[E_{3,1},E'_{3,2},E''_{3,3},\hdots,E''_{3,48}]$, where $E$, $E'$ and $E''$ represent the real, the K-SEP predicted, and the S-SEP predicted energy amounts, respectively.
  \item As all the energy amounts (or at least their estimates) are known at the beginning of the frame, use the first part of the PTF algorithm to determine the best proportional fair power allocation (sub-hours) within the frame.
  \item In the first slot of the frame, apply the power allocation found by the PTF algorithm for Slot 1 of that frame. Let, $B_{nt}=R_{nt}T$ be the number of bits that would be sent to gateway $n$ if the whole slot (of length $T$) was allocated to that gateway. If this slot is the first slot of the original frame, assign this slot to the gateway who has the maximum rate, $R_{nt}$, in that slot. Otherwise, at the beginning of each slot, $t\in\left\{2,\hdots,K\right\}$, determine the gateway with the maximum $\beta$ where, $\beta_n=\frac{B_{nt}}{\sum^{t-1}_{i=1}B_{ni}}$. Then, assign the whole slot to that gateway. If multiple gateways share the same $\beta$, then, allocate the slot to the gateway with the best channel.
 \item When first slot of the frame finishes, and thus the second slot starts, assign Slot 2 of the current frame as the first slot of the upcoming frame (half an hour shifted version of the original frame), and estimate related energy amounts. Then, add the remaining energy to the energy of the first harvest of the new frame to form a new predicted harvest series. (Ex: At 12:30, $E_{3,2}$ is known and $E_{3,3}$ can be predicted by K-SEP. The remaining 46 energy harvests are predicted by S-SEP. Thus, a new predicted harvest series is formed: $\overline{E_{pred}}=[E_{3,2}+(E_{3,1}-p_{1}T),E'_{3,3},E''_{3,4},\hdots,E''_{3,48},E''_{4,1}]$.)
 \label{it:4}
 \item Apply Step 2,3, and 4 in order until the 24 hours is completed, i.e., the last slot of the original frame has been assigned a power and time allocation.
\end{enumerate}


\section{Numerical and Simulation Results}
\label{sec:theresultsofchp7}
\subsection{K-SEP and S-SEP Related Results}
\label{sec:theresultsofchp71}
In this section, we present the numerical and simulation results related to our Kalman filter based solar energy prediction algorithm, called K-SEP, and the online resource allocation algorithm, PTF-On. By using the best weights that we computed by using the Newton algorithm, we perform numerous simulations to test our predictor. The performance of the predictor is tested by the MSE criteria and computed as follows: \begin{align}
MSE=\frac{1}{M}\sum_{i=1}^M(x_i-\widetilde{x}_i)^2
\end{align} where $x$ and $\widetilde{x}$ represent the real and estimated energies respectively, and, $M$ is the number of samples that will be considered. In order to compare the performance of our predictor with another one, we use a simple solar energy predictor, called S-SEP in this paper. S-SEP has been introduced in Section \ref{sec:ptf-on}. We first let $M=48$ (for 48 sub-hours in a day), and, compute daily MSE values for 16 days, as shown in Table \ref{tab:MSE171}. Then, average MSE over 16 days of October, 2009 (03.10-2009-18.10.2009) for K-SEP and S-SEP are, $MSE^{K-SEP}_{Aver}=4.3778 $ kilojoules/sub-hour/day and $MSE^{S-SEP}_{Aver}=84.1463 $ kilojoules/sub-hour/day respectively. By considering that the maximum power measured in~\cite{SharmaGIS10} was 60 Watts, one can produce maximum $E_{max}=60.1800=108$ kilojoules in a sub-hour by using this system. Thus, the performance of S-SEP is much worse than the performance of K-SEP in terms of average error, i.e., $\sqrt{MSE^{K-SEP}_{Aver}}=2.0923$ kilojoules/sub-hour whereas, $\sqrt{MSE^{S-SEP}_{Aver}}=9.1731$ kilojoules/sub-hour.
\begin{table*}[tb]\footnotesize
\centering
\caption{MSEs for the 16 days}
    \begin{tabular}{|l|l|l|l|l|l|l|l|l|}
    \hline
    Days         & 3          & 4        & 5        & 6        & 7        & 8       & 9        & 10      \\ \hline
    K-SEP &  0.1687    & 12.0649  &  9.5177  & 17.0761  & 6.9371   & 3.6468  & 0.3298   & 2.6919  \\ \hline
   S-SEP &  106.6895  & 266.2716 & 122.6293 &  89.9413 & 141.0086 & 43.9074 & 122.3416 & 15.8741 \\ \hline
   \hline
    Days         & 11       & 12      & 13      & 14      & 15      & 16      & 17      & 18       \\ \hline
   K-SEP &  2.0283  & 2.3286  &  3.4519 & 3.6127  &  0.2873 & 1.0939  &  2.5699 & 0.1945   \\ \hline
   S-SEP & 90.8727  & 20.1912 & 69.3250 & 26.1086 & 32.3662 &  5.6346 & 57.2630 & 135.9156 \\ \hline
    \end{tabular}
    \label{tab:MSE171}
    \vspace{-0.2in}
\end{table*}

\begin{figure}[tb]
	\vspace{-0.2in}
	\centering	
	\begin{psfrags}
		\psfrag{x}[t]{No. of sub-hours}	
		\psfrag{y}[t]{Energy (kilojoules)}
		\includegraphics[width=0.45\textwidth, height=0.35\textwidth]{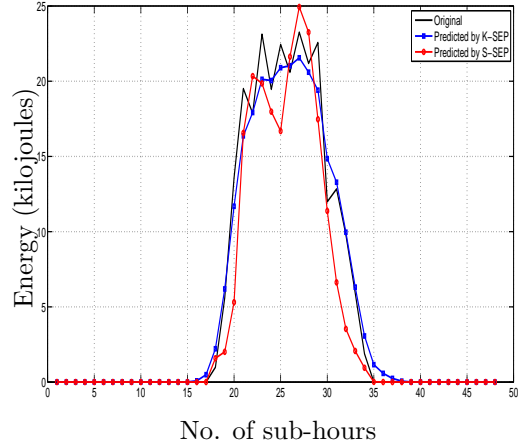}
	\end{psfrags}
	\caption{Performances of K-SEP and S-SEP compared  with the real power measurements provided in ~\cite{SharmaGIS10}; belonging to 16.10.2009 for Amherst, MA, USA.} 
	\label{fig:best}		
\end{figure}

\begin{figure}[tb]
	\vspace{-0.25in}
	\centering	
	\begin{psfrags}
		\psfrag{x}[t]{No. of sub-hours}	
		\psfrag{y}[t]{Energy (kilojoules)}
		\includegraphics[width=0.45\textwidth, height=0.35\textwidth]{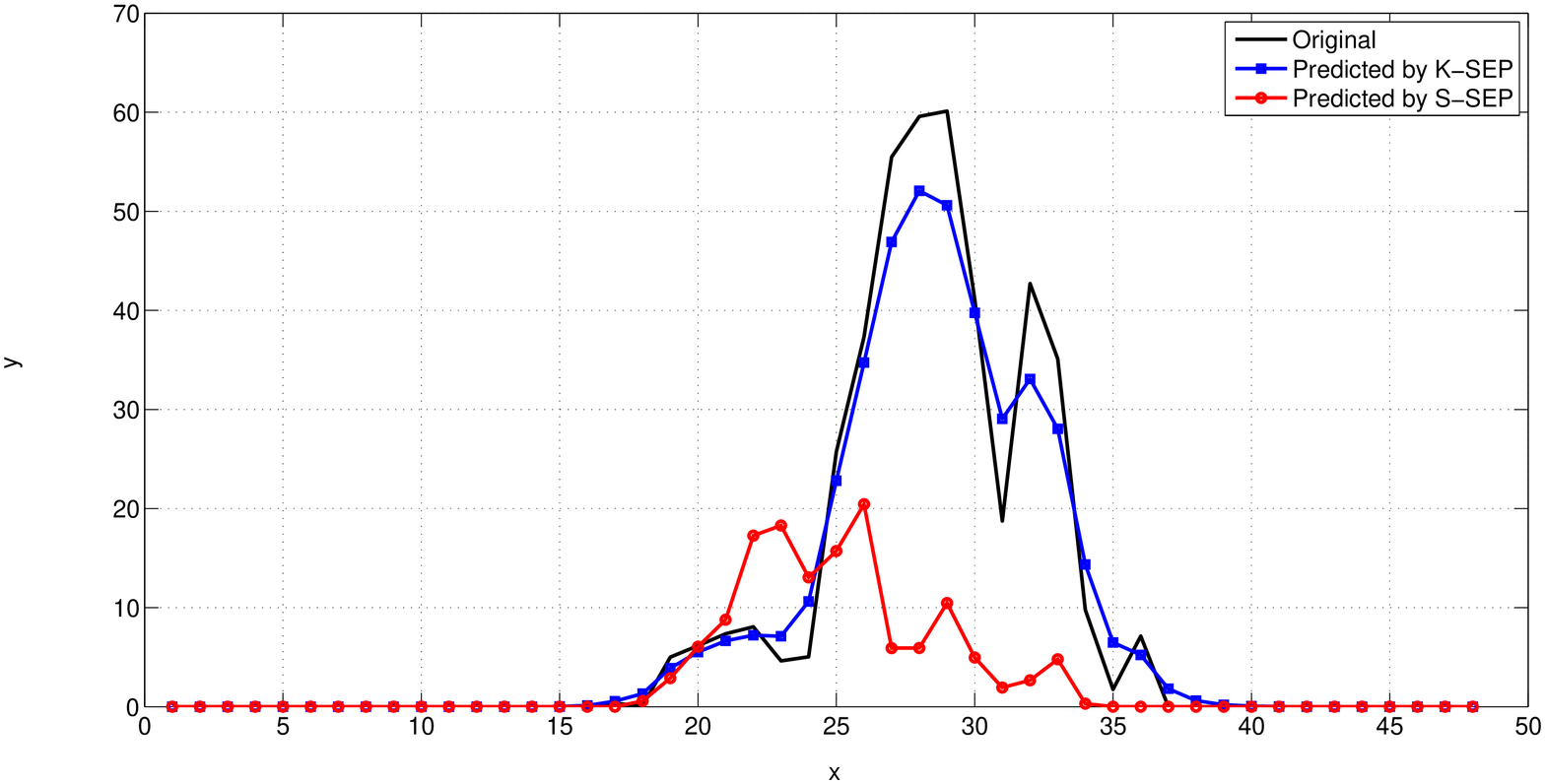}
	\end{psfrags}
	\caption{Performances of K-SEP and S-SEP compared  with the real power measurements provided in ~\cite{SharmaGIS10}; belonging to 04.10.2009 for Amherst, MA, USA.} 
	\label{fig:worst}		
\end{figure}

The figures \ref{fig:best} and \ref{fig:worst} illustrate the performances of the two predictors for two days in which S-SEP performs the best, and the worst in its 16 days's performance. As it can be seen from the figures, K-SEP outperforms S-SEP at all instances. However, even S-SEP as a simple prediction method of solar energy harvests provide some usefulness for the solution of online fashion \ref{pr:DownlinkScheduling}.     
In addition, it is more important to note that harvesting energies predicted by K-SEP algorithm always follow the original energies obtained from real measurements as shown in Figures \ref{fig:best},\ref{fig:worst}. By considering the numerical and simulation results conducted with K-SEP and S-SEP, we reach two main conclusions: the advantage of using a prediction method for the estimation of solar energy harvesting with an offline allocation algorithm (PTF) in tandem and the novelty of K-SEP which performs very close to optimal situation where energy harvests are known a priori.

\begin{figure*}[tb]
	\centering
	\begin{psfrags}
		\psfrag{number of subhours}[c]{No. of sub-hours}
		\psfrag{Energy in kJ}[b]{Energy (kilojoules)}
		\includegraphics[scale=0.35]{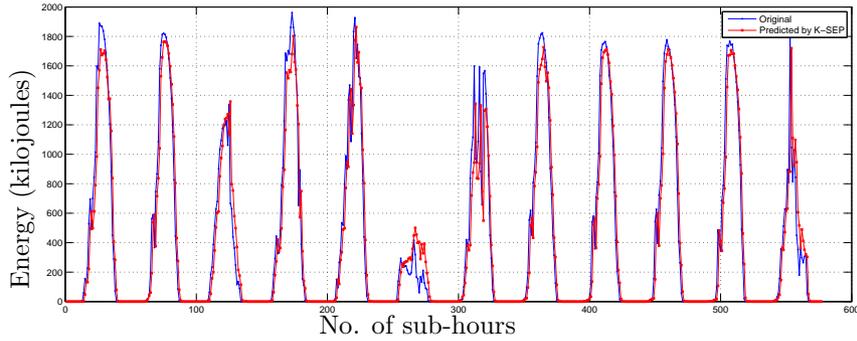}
	\end{psfrags}
	\vspace{-0.1in}
	\caption{Performance of K-SEP with respect to the real measurements from the University of Oregon Solar Radiation Laboratory; belonging to 07-19.05.2009 for Salem, MA, USA. }
	\label{fig:kalman_30min}		
\end{figure*}

Next, in order to show  the robustness and reliability of results given in Figures \ref{fig:best}, \ref{fig:worst} in Table \ref{tab:MSE171}, we further test the performance of K-SEP  with the solar irradiation measurements obtained from an entirely different source of data than \cite{SharmaGIS10}. To that end, K-SEP is run on  measurements obtained from the University of Oregon Solar Radiation Laboratory in Salem, MA, USA. Obtained results related to the performance of K-SEP algorithm can be seen from Figure \ref{fig:kalman_30min} for 12 days (between 7-19 May 2009). Similar to the previous tests, the performance of the K-SEP algorithm is very close to the original data which the all energy harvests are known in an offline manner.

\subsection{PTF-On Related Results}
\label{sec:theresultsofchp72}

In this section, we present the numerical and simulation results related to the proposed online heuristic, PTF-On. Throughout our simulations, we use the following setup: $W=10MHz$, $N_o=10^{-19} W/Hz$. For the sake of an example, we suppose that there are three sensor networks, and thus three gateways in the system, similar to the one shown in Figure \ref{fig:fig2}. The path loss of the gateways are 78, 92, and, 100 dB respectively. We compare the performance of the proposed algorithm with the performance of the ``Spend What You Get'' policy (where the amount of energy harvested at the beginning of a slot is completely spent during that slot) combined with TDMA time allocation, and with the performance of the offline PTF heuristic that is proved to operate very close-to-optimal in the authors' previous paper, \cite{6463487}.
\begin{table}[tb]\footnotesize
 \caption {FI (FAIRNESS INDEXES)’S OBTAINED BY PTF, PTF-ON, AND THE SG+TDMA SCHEME.}
 \centering
    \begin{tabular}{|l|l|l|}
    \hline
    Algotrithm & Worst Case FI & Average FI over 14 days \\ \hline
    SG+TDMA    & 0.8954                         & 0.9420                                   \\ \hline
    PTF        & 0.9345                         & 0.9531                                   \\ \hline
    PTF-ON     & 0.9084                         & 0.9299                                   \\ \hline
    \end{tabular}
 \label{tab:fair}
\end{table}
\begin{table}[tb]\footnotesize
\vspace{-0.2in}
\caption {AVERAGE OVER 14 FRAMES OF 3 GATEWAYS' THROUGHPUTS (GIGABYTES/DAY)}
\centering
    \begin{tabular}{|l|l|l|l|l|}
    \hline
    Algorithm & G1 & G2 & G3 & Total     \\ \hline
    SG+TDMA   & 188.0338  & 133.7222  & 105.8971  & 427.6531  \\ \hline
    PTF       & 458.8113  & 340.0307  & 269.5249  & 1068.3669 \\ \hline
    PTF-ON    & 473.5146  & 326.1724  & 244.9720  & 1044.6595 \\ \hline
    \end{tabular}
   \label{tab:avthr}
   \vspace{-0.2in}
\end{table}

\begin{figure*}[tb]
\vspace{-0.25in}
\centering
\begin{psfrags}
    \psfrag{No of frames}[l]{Frame no.}	
    \psfrag{Total throughput(Gigabytes)}[l]{Total throughput(Gigabytes)}
\includegraphics[scale=0.42]{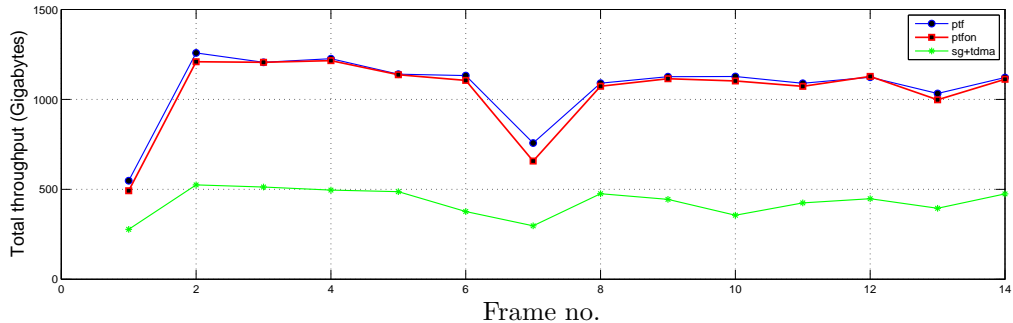}
\end{psfrags}
\vspace{-0.05in}
\caption{Total Throughput of the three gateways (in Gigabytes) over 14 Frames on Amherst, MA solar irradiation data.}
\label{fig:total}
\end{figure*}
 We start our analysis at 12:00 pm on 03.10.2009 and finish it at 12:00 on 17.10.2009. Hence, we have 14 frames, each of which consists of 24 hours (48 sub-hours). For this time period we test the performances of the PTF-On and PTF algorithms, and, the SG+TDMA scheme. The results are illustrated in Tables \ref{tab:fair}, \ref{tab:avthr} and Figure \ref{fig:total}. Note that, the $FI$ (fairness index) mentioned in Table \ref{tab:fair} is the Jain's index ($FI$), which is a well-known measure of fairness~\cite{Jain84}. $FI$ takes the value of 1 when there is a complete fair allocation, and, it is defined as $FI=\frac{(\sum^N_{i=1}x_i)^2}{N\cdot \sum^N_{i=1}x_i^2}$. For computing $FI$, we use the no. of bits transmitted to the gateways, $x_i$ for $i=1,\hdots,N$.
It is important to note that, as the utility is defined as sum of ``logarithms'' of individual throughputs, even 1\% percent improvement in utility is significant. From the viewpoint of fairness,  as illustrated in the Table \ref{tab:fair}, the performance of the proposed online algorithm, PTF-On, follows the performance of the offline PTF algorithm closely. Note that Jain's index of proposed algorithm is always higher than the threshold (0.90) by considering the worst case values during simulations. Although the average FI obtained with PTF-ON algorithm seems a little lower than the SG+TDMA algorithm; it should be remembered that the goal of PTF is to maximize proportional fairness rather than $FI$ in particular. Despite this fact, PTF-ON fares quite well with respect to FI.
\vspace{-0.02in}

Average throughputs of the three gateways obtained as a result of the simulation over 14 frames are also given in the Table \ref{tab:avthr}. As illustrated in the Figure \ref{fig:total}  and the Table \ref{tab:avthr},  PTF-ON significantly outperforms SG+TDMA. Moreover, the utility (sum of logarithms of individual rates) and the total throughput results obtained from PTF-ON are very close to the offline PTF algorithm, which was already shown to perform very close to optimal \cite{6463487}. Hence we believe that the PTF-ON algorithm, proposed as a novel solution to the power-rate allocation problem regarding proportional fairness, provides a good online solution.

\vspace{-0.1in}
\section{Conclusion}
\vspace{-0.03in}
\label{sec:conc}
This paper investigated the proportional fair power and time allocation problem in an industrial wireless sensor network system with an energy harvesting BS. The paper focuses on finding the best \textit{on-line} schedule for this problem, by predicting the energy amounts that will be harvested, and supplied to the BS, during the frame. It is proven by numerical evaluations that the joint prediction and resource allocation algorithm that we propose performs very close to the optimal offline resource allocation. The developed framework can be applied to various other scenarios , including various types of energy harvesting (such wind, vibration etc.) and various types of applications and utility functions (such as delay, reliability etc.).

\bibliography{euall_ieice}

\end{document}